\documentclass[useAMS,usenatbib,letterpaper]{mn2e}

\usepackage{natbib}
\usepackage{graphicx}
\usepackage{amssymb}

\title[Can AGN feedback-driven star formation explain the size evolution of massive galaxies? ]{Can AGN feedback-driven star formation explain the size evolution of massive galaxies? }
\author[W. Ishibashi, A. C. Fabian and R. E. A. Canning]
{W. Ishibashi$^{1}$\thanks{E-mail:
wako@ast.cam.ac.uk}, A. C. Fabian$^{1}$ and R. E. A. Canning$^{1,}$$^{2,}$$^{3}$
\footnotemark[0]\\
$^{1}$Institute of Astronomy, Madingley Road, Cambridge CB3 0HA
\footnotemark[0]\\
$^{2}$Kavli Institute for Particle Astrophysics and Cosmology, Stanford
University, 452 Lomita Mall, Stanford, CA 94305-4085, USA \\
$^{3}$Department of Physics, Stanford University, 382 Via Pueblo Mall,
Stanford, CA 94305-4060, USA}

\begin{document}

\pdfminorversion=4

\date{Accepted ? Received ?; in original form ? }

\pagerange{\pageref{firstpage}--\pageref{lastpage}} \pubyear{2012}

\maketitle

\label{firstpage}

\begin{abstract}
Observations indicate that massive galaxies at $z \sim 2$ are more compact than galaxies of comparable mass at $z \sim 0$, with effective radii evolving by a factor of $\sim 3-5$. This implies that galaxies grow significantly in size but relatively little in mass over the past $\sim$10 Gyr. Two main physical models have been proposed in order to explain the observed evolution of massive galaxies: `mergers' and `puffing-up' scenarios. 
Here we introduce another possibility, and discuss the potential role of the central active galactic nucleus (AGN) feedback on the evolution of its host galaxy. We consider triggering of star formation, due to AGN feedback, with radiation pressure on dusty gas as the driving feedback mechanism. 
In this picture, stars are formed in the feedback-driven outflow at increasingly larger radii and build up the outer regions of the host galaxy. The resulting increase in size and stellar mass can be compared with the observed growth of massive galaxies. 
Star formation in the host galaxy is likely obscured due to dust extinction and reddening. 
We suggest a number of observational predictions of our model, and discuss possible implications for AGN feedback-driven star formation.  
\end{abstract}

\begin{keywords}
galaxies: active - galaxies: evolution - stars: formation 
\end{keywords}


\section{Introduction}

Recent observations confirm that massive ($M_{\star} \sim 10^{11} M_{\odot}$) quiescent galaxies at high redshift ($z \sim 2$) are more compact than galaxies of comparable mass at $z \sim 0$ \citep{Daddi_et_2005, Trujillo_et_2007, vanDokkum_et_2008, Bezanson_et_2009, vanDokkum_et_2010, Patel_et_2012}.  
At a given stellar mass, the effective radii of compact, early-type galaxies at $z \sim 2$ can be a factor of $\sim 5$ smaller than their counterparts in the local universe \citep{vanDokkum_et_2008}. 
While massive compact galaxies seem to have been common in the past, such dense objects are found to be very rare today \citep{Trujillo_et_2009}. 
This implies that massive galaxies have undergone a substantial evolution, with a strong increase in size, over the last $\sim$10 Gyr. 
Although a certain spread in the sizes of compact galaxies is observed at high redshifts \citep{Szomoru_et_2012}, and slight biases against massive and compact objects at low redshifts cannot be excluded \citep{Taylor_et_2010}, a significant evolution in the galaxy properties is still required between $z \sim 2$ and the present. 

In particular, the central densities of the high redshift compact objects are found to be comparable to those of present-day massive ellipticals, suggesting that the main growth takes place at outer radii \citep{Bezanson_et_2009, vanDokkum_et_2010, Patel_et_2012}. 
According to the `inside-out' growth scenario, the central core of a galaxy is formed early on, while the subsequent growth occurs in the outer regions, with a gradual build-up of an extended envelope. 
A remarkable characteristic of the evolution of massive galaxies is that galaxies grow significantly in size but much less in mass. 
Indeed the evolution in effective radius is observed to be much stronger than the evolution in stellar mass. 
The combination of strong increase in size and little increase in mass provide important constraints on the possible physical mechanisms at the origin of the observed evolution.  

A number of physical models have been proposed in the literature in order to account for the evolution of massive galaxies from $z \sim 2$ to the present.  
The two main models discussed so far are the `merger' and the `puffing-up' scenarios. 
Major mergers involve mergers of galaxies with comparable mass, which lead to a linear increase in both size and mass. 
In contrast, in minor mergers the primary galaxy accretes lower mass secondary companions, and the increase in size results to be more efficient compared with the case of major mergers \citep{Naab_et_2009}.
In the puffing-up model, the size growth is obtained via a rapid mass loss from the central regions, attributed to quasar feedback \citep{Fan_et_2008, Fan_et_2010}. 

It is now widely agreed that supermassive black holes reside at the centre of all massive galaxies. 
Direct relationships between the characteristic properties of the central black hole and the host galaxy have been observed, including the correlation between black hole mass and stellar velocity dispersion and/or bulge mass \citep{Kormendy_Richstone_1995, Magorrian_et_1998, Gebhardt_et_2000, Ferrarese_Merritt_2000, Marconi_Hunt_2003, Haring_Rix_2004}. 
These empirical correlations are generally interpreted in terms of the active galactic nucleus (AGN) feedback \citep[][for a review]{Silk_Rees_1998, Fabian_1999, King_2003, Murray_et_2005, Fabian_2012}, which is expected to play a major role in the evolution of the host galaxy. 
Here we discuss the possibility of the AGN feedback triggering star formation in the host galaxy, a scenario which we have briefly explored in a previous paper \citep[][hereafter Paper I]{Ishibashi_Fabian_2012}. 
As all massive galaxies harbour massive black holes at the centre, they should have experienced some active phases during their lifetime, due to accretion onto the central black hole and associated feedback. 
However, given the actual AGN duty cycles, most galaxies in the observational samples will be detected in quiescent phases. 
In the following, we consider the active state, and try to interpret the observed evolution of massive galaxies within the feedback-triggered star formation scenario. In this paper, we explore the main concepts of the AGN feedback-driven star formation model, while the details of the model will be developed later and presented in a future work. 


\section{Feedback-triggered star formation}

We briefly summarise here the main features of our model, which is discussed in more detail in Paper I. 
We consider the effects of AGN feedback on the surrounding interstellar medium of the host galaxy, and in particular the direct effects of radiation pressure on dusty gas. Radiation from the central black hole is absorbed by dust grains embedded in the gas, and the radiative momentum is transferred to the gas through the efficient coupling between gas and dust components. 
We assume that radiation pressure on dusty gas sweeps up ambient material into a shell, and we follow the evolution of the outflowing shell in the gravitational potential of the galaxy. 
The equation of motion of the shell is given by
\begin{equation}
\frac{d}{dt} [M_g(r) \dot{r}] = \frac{L}{c} - \frac{G M_g(r) M_{DM}(r)}{r^2} \, ,  
\label{Eq_motion}
\end{equation} 
where $L$ is the luminosity of the central source, $M_g(r)$ the enclosed gas mass, and $M_{DM}(r)$ the dark matter mass. 
In Paper I, we analysed the exact conditions for the escape or trapping of the shell in the galactic halo for different underlying dark matter potentials (see also \citet{McQuillin_McLaughlin_2012}). In general, once the central black hole reaches a critical mass, or correspondingly a critical luminosity, the shell is able to expand outwards indefinitely.  

In the case of an isothermal potential, the expanding shell reaches a constant asymptotic velocity at large radii, which is independent of the initial conditions:
\begin{equation}
v_{\infty} = \sqrt{ \left( \frac{G L}{2 f_g c \sigma^2} - 2 \sigma^2 \right) } \, . 
\label{Eq_velocity}
\end{equation} 
where $\sigma$ is the velocity dispersion and $f_g$ the gas mass fraction where the gas mass is assumed to be a fraction $f_g$ of the dark matter mass. 
We may take as fiducial numerical values a typical velocity dispersion of $\sigma \sim$220 km $\mathrm{s^{-1}}$, a plausible gas mass fraction of $f_g \sim$0.1, and central luminosities of a few $L \sim 10^{47}$ erg $\mathrm{s^{-1}}$. 

The passage of the outflowing shell induces squeezing of the surrounding inhomogeneous interstellar medium. The resulting compression of the cold gas may drive gas clouds beyond their Jeans limit, inducing cloud collapse. Thus local density enhancements, caused by the passage of the outflow, may in turn lead to a triggering of star formation. 
We note that star formation may only occur in regions of enhanced density (dense clumps) within the outflow, which may have lower speeds compared to the mean shell velocity. 
The star formation rate, i.e. the global rate at which gas is turned into stars in the expanding shell, is parametrized as
\begin{equation}
\dot M_{\star} \sim \epsilon_{\star} \,  \frac{M_g(r)}{t_{flow}(r)} \, \sim \epsilon_{\star} \,  \frac{2 f_g \sigma^2}{G} v(r) \, . 
\label{Eq_SFR}
\end{equation}
where $\epsilon_{\star}$ is the star formation efficiency, and $t_{flow}(r) = \frac{r}{v(r)}$ is the local flow time. 
The observed star formation efficiency is of the order of a few percent, typically in the range $\epsilon_{\star} \sim 0.01-0.1$.
The second equality is obtained in the case of the isothermal potential in which the gas mass is directly proportional to the radius. 
We see that the star formation rate scales with the shell velocity, and is constant at large radii: $\dot M_{\star, \infty} = \dot M_{\star} (r \rightarrow \infty) \sim \epsilon_{\star} \,  \frac{2 f_g \sigma^2}{G} v_{\infty}$. 

As the shell expands outwards, new stars are formed at increasingly larger radii. 
As long as their initial velocity does not exceed the local escape velocity, newly formed stars will remain bound to the galaxy, and follow closed orbits in the galactic potential. 
We see that feedback-driven star formation mainly adds stars in the outer regions of the galaxy, thus contributing to the outer growth of the host.


\section{Size and mass growth}

New stars are successively formed in the outflowing shell according to the star formation rate given in Eq. (\ref{Eq_SFR}). 
The resulting stellar density distribution is approximately given by $M_{\star}(r) \sim \int \epsilon_{\star} \frac{M_g(r)}{r} dr$. In a simple approach, assuming that the gas density distribution can be parametrised by a power law, $\rho_g(r) = A r^{\gamma}$, the gas mass distribution follows $M_g(r) = 4 \pi \int \rho_g(r) r^2 dr  \propto r^{\gamma +3}$. 
The corresponding stellar mass profile is given by $M_{\star}(r) \propto r^{\gamma+3}$, i.e. the stellar distribution follows the underlying gas distribution with the same power law slope. 
In particular, if the gas follows an isothermal distribution, $\rho_g(r) \propto r^{-2}$, the resulting stellar distribution will also hold the characteristic isothermal profile, $\rho_{\star}(r) \propto r^{-2}$. 
A recent study suggests that the total density profiles of early-type galaxies follows a nearly isothermal profile, $\rho(r) \propto r^{\gamma}$ (with $\gamma \sim -2$), and that this particular slope can be regarded as a natural attractor \citep{Remus_et_2012}. 
Detailed future work is required to see if this applies to the model presented here. 

Once formed, stars decouple and eventually drop out from the outflowing shell. 
The driving term due to radiation pressure vanishes in the equation of motion (Eq. \ref{Eq_motion}), and we are left with only the gravitational term. 
Thus the newly formed stars start to follow their own ballistic orbits in the gravitational potential of the galaxy. 
For simplicity, we may assume that stars initially share the properties of the expanding shell in which they are formed, with the initial velocity corresponding to the local shell velocity at the formation radius. 
The initial conditions are given by: $r_{\star,0} = r_{\mathrm{shell}}$ and $v_{\star,0} = v_{\mathrm{shell}}(r_{\star,0})$. 
In a first stage, stars continue to move outwards following their initial radial motion.
But under the effects of gravity, they will soon be slowed down, and later fall back, accelerating towards the centre. 
As a result, stars orbit back and forth in the gravitational potential of the galaxy, and on average spend most of their time where their velocity is lower. 
For simplicity, we may also assume that the star's trajectory is symmetric with respect to the centre. 
We can then define the mean or average radius of the orbiting star as
\begin{equation}
\bar r = \frac{\int r(t) dt}{\int dt} \, . 
\end{equation} 
In the isothermal potential, the resulting average radius can be written as a multiple of the initial radius: $\bar r = K \times r_{\star,0}$, where the normalisation constant $K$ depends on the initial velocity. 
For initial velocities in the range $\sigma < v_{\infty} < 2 \sigma$, we obtain that the corresponding mean radius lies in the range $r_{\star,0} < \bar r < 2r_{\star,0}$. 
Evidently, a high initial velocity leads to a larger value of $K$, and hence a larger mean radius, up to a doubling of the initial radius. 
Further, a high initial velocity also implies that the expanding shell had already travelled to greater distances. 
In fact the maximal radius reached by the outflowing shell within a given AGN activity timescale is roughly given by: $r_{max} = r_0 + v_{\infty} \cdot \Delta t_{AGN}  \sim v_{\infty} \cdot \Delta t_{AGN}$, where $r_0 \ll v_{\infty} \cdot \Delta t_{AGN}$.  
Thus a high shell velocity implies both a large initial radius and large initial velocity, which couple to give a large mean radius for the orbit of the star. 
The formation of new stars in the outer regions can therefore lead to a substantial increase in the galaxy size. 

In physical terms, the increase in shell velocity, which ultimately leads to the increase in galaxy size, can be linked to the decrease of the gas mass fraction over cosmic time. 
Indeed high-redshift ($z \sim 2$) galaxies are observed to be much more gas-rich compared to present-day galaxies. Measurements of the baryonic gas fraction ($f_g^{'}$, defined as the ratio of gas mass to the sum of gas and stellar masses), yield values of the order of $\sim 0.5$ in normal star-forming galaxies, close to the peak epoch of galaxy assembly activity \citep{Daddi_et_2010, Tacconi_et_2010, Tacconi_et_2012}. These observed gas fractions are at least a few times higher than the typical values measured in local counterparts, suggesting a significant evolution in the gas mass fraction. 
The gas phase thus accounts for a major fraction of the baryonic mass in high-redshift galaxies, with gas masses being comparable to stellar masses. 
The large molecular gas content of these gas-dominated galaxies has been suggested as a possible explanation for the higher specific star formation rates observed at high redshifts \citep{Tacconi_et_2010}. 
Interestingly, new Hubble Space Telescope (HST) observations reveal the presence of equally significant amounts of cool gas in the haloes of early-type galaxies \citep{Thom_et_2012}. 
The detection of cold, bound gas in early-type galaxies seems to be comparable to that of star-forming galaxies. 
This implies that passive galaxies also contain important reservoirs of cold circumgalactic gas, which can potentially be re-accreted and eventually made available for star formation. 
Therefore the quiescence of $z \sim 2$ galaxies, and associated quenching of star formation, may not be entirely attributed to the lack of cold gas reservoir. 

A rough order of magnitude estimate of the total stellar mass added by the feedback-driven star formation process can be obtained by integrating the star formation rate (Eq. \ref{Eq_SFR}) over the AGN feedback timescale. 
Assuming radiatively efficient accretion, we may take the Salpeter time as the characteristic feedback timescale. 
We obtain typical star formation rates of $\gtrsim$100 $M_{\odot}$ $\mathrm{yr^{-1}}$, and associated stellar masses of the order of $\sim 10^{10} M_{\odot}$ per cycle. 
We note that the increase in mass due to the feedback-driven star formation is not huge, but still a non negligible fraction of the total stellar mass. 
Thus a few such episodes occurring during the lifetime of the galaxy could contribute to the overall stellar mass growth. 

In the central regions, the radiation field can be very strong and efficiently heat the surrounding medium via photoionization and Compton scattering. In particular, Compton heating may dominate in the innermost regions, where matter is highly ionized. 
Such a Compton-heated zone can extend up to a distance of $\sim$1 kpc for a typical quasar spectral energy distribution \citep{Sazonov_et_2005}. 
Thus gas is prevented from cooling, and consequently star formation is inhibited in this central region. 
Compton heating may therefore be responsible for quenching star formation in the inner core of the host galaxy. 
Indeed numerical simulations, including radiative feedback, show the presence of an inner $\sim$kpc region heated to high temperatures via Compton scattering, where the amount of cold gas is significantly reduced and star formation is locally suppressed \citep{Kim_et_2011}.


\section{Some observational predictions}

Here we set out a number of observational predictions of our model.
We have seen that stars formed in the outflowing shell eventually follow bound orbits in the gravitational potential of the galaxy. The mean radius of the stars is mainly determined by the initial conditions: in general, high shell velocities lead to large mean radii. As a result, a significant increase in the size of the galaxy can be easily obtained.  
On the other hand, the associated stellar mass added in the process is not very large. 
Considering a few AGN feedback episodes during the lifetime of a galaxy, we would thus expect a strong increase in size but with a modest increase in stellar mass. 
This characteristic property can be directly compared with the observed size and mass evolution of massive galaxies. 
We should recall, however, that our model describes active phases when the central AGN is switched on, while galaxies selected in observational samples are mostly found in the quiescent regime. 

In our picture, star formation is triggered by AGN feedback and new stars are formed in the feedback-driven shell. 
We therefore expect a good correlation between star formation and outflow properties. 
We predict the star formation rate to scale with the outflow velocity (Eq. \ref{Eq_SFR}); we also expect more luminous, or equivalently more massive, objects to have higher star formation rates. 
The newly formed stars are likely to follow radial orbits, at least in the initial phases. 
Outflows are now commonly observed in galaxies, and the expected connection with star formation activity can be directly searched for in the observational data. 

Star formation in the outflowing shell can be traced by UV and H$\alpha$ emissions, characteristic of young stellar populations. 
In cases where the H$\alpha$ emission can be directly detected, a detailed analysis of its emission line profiles and in particular broad wings, can give indications on the radial motion of the stars. 
However, much of the intrinsic UV emission associated with star formation could be absorbed by dust grains, resulting in obscuration and reddening.  
We recall that radiation pressure on dust is the driving feedback mechanism, which pushes material out to large radii, and thus expect significant amounts of dusty gas spread over the entire galaxy. 
We also note that the formation of new stars is expected to occur in dense clumps within the outflowing shell, which may travel more slowly compared to the mean shell speed at any radius. 
This suggests that dusty gas might travel ahead of the stars, and form a sort of screen shielding the intrinsic emission of young stars. 
We would therefore predict obscured star formation and dust reddening in massive growing galaxies. 

As new stars are formed at ever larger radii, a low-density envelope builds up in the outer regions of the host galaxy. 
The galaxy outskirts are likely to have bluer colours, due to the presence of young stellar populations, compared to the inner core, where Compton heating inhibits star formation. 
Negative colour gradients are then expected in the radial direction, with a central red core and bluer outskirts. 
But radial colour gradients might simply reflect metallicity gradients, and one needs to disentangle the two effects. 
If the outflowing shell remains trapped in the galactic halo, gas may eventually fall back and provide fuel for further accretion and star formation in a subsequent episode. 
New HST observations indeed suggest that plenty of cool gas, potentially available for re-accretion, is bound to the haloes of early type galaxies \citep{Thom_et_2012}. 
In a cycling scenario, material can be continuously recycled, leading to metal enrichment. 
Subsequent feedback cycles are fuelled by recycled gas which failed to undergo star formation and has fallen back in as well as more pristine gas from cold flows of external gas.
This, in turn, could lead to a complex relationship between metallicity and feedback through the dust component, which is the underlying driving mechanism. 

Due to the entwinement between star formation and dust absorption, the detailed colour evolution of a galaxy is complex and difficult to predict. New stars are formed in the feedback-driven shell, and at the same time dusty gas is spread out to large radii. We thus expect intrinsically bluer colours in the outskirts due to the presence of young stellar populations, but also reddening effects due to the presence of dust. The dust and stellar evolutions are closely coupled, essentially occurring on comparable timescales, and the resulting colour evolution is time-dependent. 

In the above sketched picture, we have implicitly considered the simple case of an isolated galaxy. 
But in a more realistic situation, the galaxy is likely to reside in a group or cluster environment, where tidal interactions with neighbouring objects can be important. 
In particular, the extended low density envelopes of the growing galaxies can be easily affected by tidal torques, and possibly disrupted or even induced to assume some form of rotational motion by the passage of another galaxy in the vicinity. 
Furthermore, if the outflow has a bipolar pattern, as expected if radiation is originating from an inner accretion disc, then the growing galaxy might present an elongated shape, which may superficially resemble a disc-like object. 
This may in part account for the disk-dominated morphologies reported in recent observations of high-redshift ($z \sim 2$) galaxies \citep{vanderWel_et_2011}, as it could be difficult to clearly distinguish the observed velocity field from rotational motion.  

Finally, since our model is based on AGN feedback, we expect the feedback-driven star formation to be most effective at high redshifts, close to the peak epoch of AGN activity ($z \sim 2$) \citep{Merloni_Heinz_2008}. 
Therefore this particular mode of star formation should mainly contribute to the galaxy growth in the high redshift regime ($1 \lesssim z \lesssim 2$). 
In this context, it is interesting to note that the peak epoch of both AGN and star formation activities are found to be roughly coincident, and it is even possible that this parallel evolution is triggered by AGN feedback.


\section{Discussion}

In our picture, star formation is triggered by AGN feedback and takes place in the feedback-driven shell. 
We thus expect correlations between star formation and outflow properties. 
There is now ample observational evidence of galactic-scale outflows detected in several sources \citep[][and references therein]{Sturm_et_2011, Tombesi_et_2012}. 
However, contrasting results concerning possible connections between AGN and star formation activities have been reported in the literature \citep{Page_et_2012, Santini_et_2012, Rovilos_et_2012, Harrison_et_2012}. 
The relation is usually analysed by trying to separate the dominant emission components: X-rays are assumed to trace AGN activity, while far infrared/sub-millimetre emission is considered as a proxy for star formation. 

Based on Herschel SPIRE sub-millimetre observations, \citet{Page_et_2012} claim that star formation is suppressed in powerful AGNs with X-ray luminosities exceeding $L_X > 10^{44}$ erg $\mathrm{s}^{-1}$ in the redshift range $1 < z < 3$. 
In contrast, \citet{Santini_et_2012}, using Herschel PACS observations, find that AGN hosts have enhanced star formation with respect to inactive galaxies of comparable mass in the redshift range $0.5 < z < 2.5$. 
Further, dividing the sample into high- and low- X-ray luminosities, they observe that the star formation enhancement is more important in more luminous objects. 
In a recent analysis, combining deep XMM-Newton and Herschel observations, a positive correlation is found between the specific star formation rate (sSFR) and X-ray luminosity for luminous objects ($L_X \gtrsim 10^{43}$ erg $\mathrm{s}^{-1}$) at high redshifts ($z \gtrsim 1$); whereas no such correlation is detected in lower luminosity and lower redshift systems \citep{Rovilos_et_2012}. 
Another recent study, extending the analysis to a larger sample including the COSMOS field, show that there is no clear evidence of suppression of star formation in high-luminosity AGNs at z$\sim$1-3 \citep{Harrison_et_2012}. Instead, the mean star formation rate of luminous objects ($L_X > 10^{44}$ erg $\mathrm{s}^{-1}$) is found to be comparable to that of lower luminosity sources, also consistent with the star formation rates of typical star-forming galaxies at similar redshifts ($\langle \mathrm{SFR}\rangle$ $\sim$100 $M_{\odot}$$\mathrm{yr^{-1}}$). Although the issue is still under debate, there seems to be some observational evidence for a correlation between AGN and star formation activities.
Such positive correlations are expected in the initial phases of the feedback-driven star formation and would be naturally explained in our framework. 

A particular characteristic emerging from observational studies of the evolution of massive galaxies is the inside-out growth pattern. 
The huge masses coupled with the small radii observed in the high-redshift compact galaxies imply extremely high densities. 
But the central densities (measured within a fixed physical radius of $\sim$1 kpc) of massive galaxies at $z \sim 2.3$ are found to be quite similar to that of local elliptical counterparts, leading to the suggestion that the compact galaxies may form the core of present-day ellipticals \citep{Bezanson_et_2009}. 
Analysis of massive galaxies, selected at constant number density in the redshift range $0 < z < 2$, also show that the mass within a fixed inner radius of 5 kpc is almost constant, while the main mass growth takes place at larger radii \citep{vanDokkum_et_2010}. More recent results, based on Hubble Space Telescope (HST) imaging, confirm that most of the mass within the central 2 kpc is already assembled at $z \sim 2$, and that the subsequent growth occurs in the outer regions of the galaxy \citep{Patel_et_2012}. 
All these observations seem to indicate that massive galaxies grow inside-out, gradually assembling their stellar mass around the already formed compact cores. 

Another important feature is that the evolution in size and mass are not parallel: the increase in size is much stronger than the increase in mass. The relation between the effective radius and the stellar mass can be parametrized as $R_e \propto M^{\alpha}$ with $\alpha \gtrsim 2$ \citep{Patel_et_2012}.
In addition, the bulk of the mass and size evolution seem to occur at $z \gtrsim 1$ \citep{Perez-Gonzalez_et_2008, Fan_et_2010}, with the growth rate being faster at higher redshifts \citep{Newman_et_2012}.  

As mentioned in the Introduction, different physical mechanisms have been proposed in order to explain the observed size and mass evolution of massive galaxies. 
In the puffing-up scenario, quasar feedback removes large amounts of gas from the central regions, leading to an expansion of the system at fixed stellar mass \citep{Fan_et_2008, Fan_et_2010}. 
As there is no increase in mass, while the size increases significantly, the velocity dispersion of the galaxy is expected to decrease considerably. 
In contrast, observational measurements indicate only a mild evolution in the velocity dispersion \citep{Cenarro_Trujillo_2009}. 
The puffing-up scenario is therefore not supported by observations.

In major mergers of two equal mass galaxies, the conservation of kinetic energy implies $R_e \propto M$, and the velocity dispersion remains constant. 
This indicates a linear increase in both size and mass, e.g. a doubling in size requires a doubling in mass. 
If the observed size growth is attributed to major mergers, the corresponding mass growth would yield too many massive galaxies ($\sim 10^{12} M_\odot$), in contradiction with the number density observed in the present day universe \citep{Bezanson_et_2009, Hopkins_et_2010}. 
Moreover, direct estimates of the pair fraction of massive galaxies suggest that the merger fraction is not sufficient to account for the observed size growth from $z \sim 2$ to $z \sim 0$ \citep{Man_et_2012}. 
Therefore major merging can be ruled out as the main growth mechanism for massive galaxies. 

In the case of minor mergers, the primary galaxy accretes material by merging with lower mass companions (typically with ratio 1:10). Conservation of energy implies that the effective radius is proportional to the square of the mass, $R_e \propto M^2$. 
Thus for a given increase in mass, the corresponding increase in size is much more efficient compared to the case of major mergers. 
Numerical simulations of minor mergers also show that the accreted low-density material tends to settle in the outer parts of the primary galaxy, implying a growth pattern consistent with the inside-out growth scenario \citep{Naab_et_2009, Oser_et_2012}. 
Therefore the minor merger model is currently considered as the favoured interpretation in accounting for the observed evolution of massive galaxies \citep{Bezanson_et_2009, Hopkins_et_2010}.
However, recent results from the CANDELS survey suggest that minor mergers may only explain the size growth at z $\lesssim$ 1; whereas the faster growth rate observed at higher redshifts cannot be accounted for, and requires additional physical processes \citep{Newman_et_2012}. 

In our model, new stars are formed in the AGN feedback-driven shell at increasingly larger radii and populate the outer regions of the host galaxy. 
It is also possible that some of the stars will be pushed beyond the gravitational potential of the host and be ejected from the galaxy. Such young stars distributed in the halo may form part of the intergalactic stellar population and contribute to the enrichment of the intracluster medium. 
We have seen that the increase in galaxy size is accompanied by a modest increase in stellar mass, in agreement with the observational trend. 
The succession of several such feedback-driven star formation episodes may thus account for the evolution of massive galaxies. 
Most of the galaxy growth is observed to occur at large radii, while the central region remains roughly constant \citep{Bezanson_et_2009, vanDokkum_et_2010, Patel_et_2012}. 
The absence of significant star formation in the central core, due to Compton heating, may explain the similarity of the core component observed in massive galaxies. 

Massive, compact galaxies at $z \sim 2$ are often observed in a quiescent phase. Deep spectroscopic observations of a typical compact quiescent galaxy at $z \sim 2.2$ suggest that quenching of star formation can be very effective at these redshifts \citep{Kriek_et_2009}. As a result, a red-sequence is likely to be already in place at $z \sim 2$, with galaxies mainly in a post-starburst phase. However, the colour evolution of the red-sequence to the present cannot be explained by just assuming passive evolution: simple aging of the stellar populations would produce galaxies at $zÊ\sim 0$ that are redder than observed \citep{Kriek_et_2008}. New starburst events occurring in red-sequence galaxies at later times might help in reducing the strong colour evolution. 
In our picture, several episodes of star formation associated with the AGN feedback activity are envisaged over the lifetime of a galaxy. 
As young stars are formed at increasingly larger radii in the outflowing shell, we also expect a radial colour gradient to develop within the galaxy, leading to bluer outskirts. 
Indeed negative colour gradients are observed in massive early type galaxies in the redshift range $1.3 < z < 2.5$, with colours becoming bluer with increasing radius \citep{Guo_et_2011}. 
Similar negative colour gradients in the radial direction are detected in a sample of galaxies at $z \sim 2$ with a variety of morphologies \citep{Szomoru_et_2011}. 

Compact galaxies at high redshifts are observed to be `red and dead', with little gas and no ongoing star formation. 
It is plausible that an earlier episode of AGN feedback has removed cold gas from the core of the galaxy, but gas should still be present in the outskirts of the galaxy and in the surrounding circumgalactic medium, trapped in the dark halo.  
Indeed large reservoirs of cool gas are observed in the haloes of early-type galaxies at $z < 1$ \citep{Thom_et_2012}, and recently a significant amount of cool gas extending to hundreds of kiloparsecs has also been detected in quasar host haloes at $z \sim 2$ \citep{Prochaska_et_2013}. 
Over the cosmic time, from $z \sim 2$ to the present, we expect galaxies to undergo several gas re-accretion phases, which should power AGN activity and associated feedback-driven star formation cycles. 
The observed reservoirs of gas surrounding the progenitors of today's ellipticals may provide the required gas fuel. 

A significant amount of dust ($A_V \gtrsim 1$) is also required in order to efficiently drive AGN feedback. 
Recent results based on Herschel observations of the Crab Nebula suggest that the production of dust in core-collapse supernovae can be quite efficient \citep{Gomez_et_2012}. This implies that new dust can be produced by supernova explosions and spread into the surrounding star-forming medium, enhancing the overall feedback process. 
At the same time, dust may also be responsible for extinction and reddening. 
We thus expect dust-obscured star formation in the host galaxy. 
The observed red colours of massive galaxies at high redshifts are usually interpreted as due to old stellar populations, with negligible star formation. But dust reddening may also play a non negligible role, and the observed red colours might be partially attributed to dust effects.
In fact, samples of galaxies at $1.5 < z < 3$ selected from the GOODS-NICMOS survey show that massive galaxies have high star formation rates, of the order of $\sim$100 $M_{\odot}$$\mathrm{yr^{-1}}$, but are still dominated by red rest-frame (U-B) colours \citep{Bauer_et_2011}.
This suggests that star formation in most of these systems is obscured by dust. 
A number of heavily reddened quasars, requiring large dust extinction ($A_V \sim 3$) have also been recently observed at $z \sim 2$ \citep{Banerji_et_2012}. 
Thus there seems to be some observational evidence for dusty star formation and dust reddening occurring in massive galaxies at high redshifts. 

Different forms of star formation triggering have been previously discussed in other contexts.
Early studies suggested triggered star formation, associated with the propagation of relativistic jets, as a possible explanation for the observed alignment of optical and radio emissions in radio galaxies \citep{Rees_1989}. 
A jet-triggered positive feedback can also enhance the star formation in luminous starbursts \citep{Silk_2005}. 
\citet{Silk_Nusser_2010} invoke star formation triggered by AGN winds to boost feedback, from the combination of both AGN and supernovae, in order to drive gas out of the host galaxy.  
More recently, detailed numerical simulations of radio mode feedback, with jet-induced star formation, have also been performed \citep{Gaibler_et_2012}. 
While jet-triggered star formation is mainly associated with radio-loud objects, the radiation pressure-driven feedback considered here may be applied to a wider range of sources. 

The particular galaxy growth mechanism discussed here mainly operates at high redshifts, where AGN activity is more pronounced, and where the rate of minor mergers seem to be insufficient. 
Therefore AGN feedback-driven star formation may explain the size growth at high redshifts ($1 \lesssim z \lesssim 2$), while minor merging may play a major role at lower redshifts ($0 \lesssim z \lesssim 1$), with a combination of different physical mechanisms dominating at different epochs.


\section*{Acknowledgments}

We thank Richard Ellis for helpful comments and Roberto Maiolino for discussions. 
WI acknowledges support from the Swiss National Science Foundation.

\bibliographystyle{mn2e}
\bibliography{biblio.bib}


\label{lastpage}

\end{document}